\documentclass[pra,letterpaper,superscriptaddress,twocolumn]{revtex4}%

\usepackage{amsfonts}
\usepackage{amsmath}        
\usepackage{amssymb}
\usepackage{graphicx}
\usepackage{bbold}
\usepackage[usenames]{color}
\usepackage{tikz}%
\providecommand{\U}[1]{\protect\rule{.1in}{.1in}}

\begin{document}
\title{Scattering of two particles in a 1D lattice}
\author{Seth T. Rittenhouse\footnote{corresponding author: rittenho@usna.edu}}
\affiliation{Department of Physics, the United States Naval Academy, Annapolis Maryland
21402, USA}
\author{P. Giannakeas}
\affiliation{Max-Planck-Institut für Physik komplexer Systeme, N\"{o}thnitzer Street 38, D-01187 Dresden, Germany}
\author{Nirav P. Mehta}
\affiliation{Department of Physics and Astronomy, Trinity University, San Antonio, Texas
78212, USA}

\begin{abstract}
This study concerns the two-body scattering of particles in a one-dimensional periodic potential. A convenient ansatz allows for the separation of center-of-mass and relative motion, leading to a discrete Schr\"odinger equation in the relative motion that resembles a tight-binding model.  A lattice Green's function is used to develop the Lippmann-Schwinger equation, and ultimately derive a multi-band scattering K-matrix which is described in detail in the two-band approximation. Two distinct scattering lengths are defined according the limits of zero relative quasi-momentum at the top and bottom edges of the two-body collision band. Scattering resonances occur in the collision band when the energy is coincident with a bound state attached to another higher or lower band. Notably, repulsive on-site interactions in an energetically closed lower band lead to collision resonances in an excited band.
\end{abstract}
\date{\today}
\maketitle

\section{Introduction}
Ultracold gases embedded in optical lattices present numerous theoretical and experimental opportunities for the investigation of few- and many-body physics \cite{bloch_NP_2005,bloch_RMP_2008}.
Such systems provide a versatile platform for a number of reasons. Control of the laser intensity, wavelength and beam geometry enable  detailed tunability of the depth, spacing, and geometry of the lattice. Moreover, the variety of atomic species that have been successfully trapped includes bose, fermi and even mixed-symmetry systems~\cite{bloch_RMP_2008,gunter2006PRL,lewenstein2004PRL}, all of which can be studied by tuning their mutual interactions via Feshbach resonances~\cite{chinFeshbachResonancesUltracold2010}.
For example, bosonic ensembles in a lattice permitted the realization of a many-body phase transition from superfluid to Mott insulator \cite{greinerQuantumPhaseTransition2002,fisherBosonLocalizationSuperfluidinsulator1989}, and site-resolved imaging of Bose~\cite{bakr2010Science,sherson2010Nature} and Fermi~\cite{greif2016Science,cheuk2016PRL} systems has enabled yet more flexibility.

In addition, two-dimensional (2D) Fermi gases in lattices 
are proposed as candidates to study topological many-body phases such as $p$-wave superfluidity \cite{fedorovpwavepra2017}. 
Recently, ultracold atoms in driven optical lattices proved to be a panacea for the experimental realization of time crystals \cite{auttiACJosephsonEffect2021,choiObservationDiscreteTimecrystalline2017,zhangObservationDiscreteTime2017}; an exotic many-body phase that features a broken translation symmetry both in space and time, where Wilczek's \cite{wilczekPRL2012} initial proposal laid the ground for a more systematic theoretical understanding \cite{sacha2017time,sachapra2015,lazaridesPRL2014,lazaridesPRL2015}.
Furthermore, non-equilibrium dynamics in 1D lattices induced via interaction quenches on few-bosonic ensembles result in the formation of global density-waves \cite{mistakidis2014interaction,mistakidisPRA2015}, directional transport by spatially modulated interactions \cite{Pla_mann_2018}, and many-body expansion in weakly interacting Bose-Fermi mixtures \cite{sieglPRA2018}.

Apart from these advances in the realm of many-body physics, studies on the few-body aspects of ultracold atoms in lattice geometries explore their multi-faceted collisional properties, such as the formation of bound pairs \cite{winklerRepulsivelyBoundAtom2006,valienteTwoparticleStatesHubbard2008,Nygaard2008PRA,petrosyanQuantumLiquidRepulsively2007,piilTunnelingCouplingsDiscrete2007,gruppResonantFeshbachScattering2007}, lattice-induced resonances \cite{fedichevExtendedMoleculesGeometric2004, vonStecher2011PRL, Wouters2006PRA}, Feshbach resonances in lattices \cite{dickerscheidFeshbachResonancesOptical2005}, and the physics of reactive and Umklapp processes \cite{TerrierPRA2016}.
More refined theoretical studies on the two-body collisional physics permitted the inclusion of finite range effects \cite{Valiente2010PRA} and explored the impact of the energetically higher-bands \cite{TerrierPRA2016,Cui2010PRL}.
Two-body collisions on a lattice occur within a set of energy bands which loosely behave as collision channels, with two-body interactions yielding intra- and inter-band effects on collisional processes \cite{Cui2010PRL} similar to the behavior seen in confinement induced resonances \cite{bergeman_etal}.
Beyond the two-body physics, theoretical studies have shown the existence of three-body bound states in three-dimensional and 1D lattices \cite{mattisFewbodyProblemLattice1986,valienteThreebodyBoundStates2010a}.
Additionally, in such systems an on-site attractive three-body interaction can emerge that induces an instability yielding thus the collapse of the many-body ground state \cite{paulHubbardModelUltracold2016}.  Evidently, the detailed understanding of scattering processes in lattice geometries, and the necessary conditions under which resonant phenomena can occur is of paramount importance to the design and manipulation of exotic many-body phases.

In this work a systematic pathway to address collisional physics of two particles, with either bosonic or fermionic character, in the presence of a periodic potential is developed based on the $K-$matrix formalism.
Within this formalism, the energy-normalized Bloch states are employed as scattering waves incorporating contributions from both energetically accessible (open) and inaccessible (closed) bands.
In agreement with previous works, we observe that for attractive interactions comparable in strength to the band gap, a resonance arises for scattering between particles in the lowest band due to virtual transitions into the closed bands \cite{Valiente2010PRA,Cui2010PRL,Nygaard2008PRA,vonStecher2011PRL,Orso2005PRL,Wouters2006PRA}
Furthermore, we observe that for repulsive on-site interactions with energy similar to the band gap additional resonant features occur for scattering of two particles in an \emph{excited} band.

The paper is organized as follows: In Sec.~\ref{section:1Dlattice}, we describe the two-body,
multi-band Hamiltonian in a basis of Wannier states, and separate it into
discrete center-of-mass and relative motion coordinates. In Sec.~\ref{section:latticeGF} we
develop the Green's operator for the relative non-interacting Hamiltonian via
a lattice Green's function for both energetically open and closed bands. In
Sec.~\ref{section:scattering} we derive the lattice K-matrix for two-body, on-site interactions,
examine the special case where each particle can occupy only 2 energy bands.
In Sec.~\ref{section:scatlen} we derive the lattice scattering length which can be used to
describe the interaction between two particles at energies near the top or
bottom of the two-body bands. Finally in Sec.~\ref{section:summary} we summarize our results and
discuss future work.

\section{1D lattice}
\label{section:1Dlattice}

In this section, we describe a system consisting of two particles
confined to a periodic one-dimensional (1D) potential $V_{\text{lat}}\left(  x\right)  $ with
periodicity $\lambda$. In the absence of a two-body interaction the behavior of each
particle is described by the simple Hamiltonian $\widehat{\mathcal{H}}
_{0}=-\tfrac{\hbar^{2}}{2m}  \tfrac{d^{2}}{dx^{2}}+V_{\text{lat}}\left(
x\right)  $ which is diagonalized by a Bloch function $\phi_{\mu}\left(
q;x\right)  $
where $\mu$ is a band index and $q$ is the quasi-momentum. Bloch
waves are delocalized functions that extend throughout the entire lattice.
However, they can be combined into an orthonormal basis localized to each
lattice site for each band. These are the Wannier functions which take the form%
\begin{equation}
w_{\mu n}\left(  x\right)  =\left(  \dfrac{\lambda}{2\pi}\right)  ^{1/2}%
\int_{-\pi/\lambda}^{\pi/\lambda}e^{iqn\lambda}\phi_{\mu}\left(  q;x\right)
dq. \label{Eq:wannierdef}%
\end{equation}
Here, beyond the band index $\mu,$ there is an additional index in the Wannier
functions; the site index, $n\in\mathbb{Z}$, specifying the lattice site
location $x=n\lambda$ at which the function is localized.

The behavior of a single particle in the lattice is characterized by the
Hamiltonian%
\begin{align}
\hat{h}= & \sum_{\mu,n}\epsilon_{\mu}\left\vert \mu n\right\rangle
\left\langle \mu n\right\vert \label{Eq:onebodham} \\ 
 & -\sum_{\mu,n,j>0}S_{j}^{\mu}\left(  \left\vert
\mu n\right\rangle \left\langle \mu\left(  n+j\right)  \right\vert +\left\vert
\mu\left(  n+j\right)  \right\rangle \left\langle \mu n\right\vert \right)  \nonumber,
\end{align}
where $\left\vert \mu n\right\rangle $ is the Wannier state associated with a
particle in the $\mu$th band localized to the $n$th lattice site. Here,
$\epsilon_{\mu}=\left\langle \mu n\left\vert \widehat{\mathcal{H}}%
_{0}\right\vert \mu n\right\rangle $ is the onsite energy and $S_{j}^{\mu
}=-\left\langle \mu n\left\vert \widehat{\mathcal{H}}_{0}\right\vert
\mu\left(  n+j\right)  \right\rangle $ is the energy associated with the
particle hopping $j$ sites from site $n$ to site $n\pm j$. 
Note that we have
assumed that the band energies are symmetric in the quasi momentum $q$.
Diagonalizing this Hamiltonian gives, not at all surprisingly, the band
dispersion relation written as its cosine Fourier transform, i.e.%
\begin{equation}
E_{\mu}\left(  q\right)  =\epsilon_{\mu}-\sum_{j=1}^{\infty}2S_{j}^{\mu}%
\cos\left(  jq\right)  . \label{Eq:dispersion1B}%
\end{equation}
For localized Wannier functions we can expect that tunneling to more distant sites will be suppressed. This results in a strong suppression in the hopping energy $S_{j}^{\mu}$ for $j>1$. Thus, for the purposes of this work we
will assume only nearest neighbor hopping terms survive, i.e. $S_{j}^{\mu
}=S_{\mu}\delta_{1,j}$. 

With the single particle discrete Hamiltonian in hand, we may now proceed to write the two-body Hamiltonian in terms of the localized discrete Wannier basis. In first quantized form, the full Hamiltonian is given by%
\begin{equation}
\hat{H}=\hat{h}_{1}+\hat{h}_{2}+\hat{V} \label{Eq:fullham}%
\end{equation}
where $\hat{h}_{j}$ is the single particle Hamiltonian Eq.~(\ref{Eq:onebodham})
for particle $j$, and $\hat{V}$ is the interaction between the two particles.
In the Wannier basis, the interaction is expressed as
\begin{equation}
\hat{V}=\sum_{\substack{\mu,\nu,\mu^{\prime},\nu^{\prime}\\m,n}}U%
_{\mu^{\prime},\nu^{\prime}}^{\mu,\nu}\left(  \left\vert n-m\right\vert
\right)  \left\vert \mu m;\nu n\right\rangle \left\langle \mu^{\prime}%
m;\nu^{\prime}n\right\vert , \label{Eq:intham}%
\end{equation}
Here, $\left\vert \mu m;\nu n\right\rangle $ represents the two-body Wannier state of particle 1 in band $\mu$ localized to site $m$ and particle 2 in band
$\nu$ localized to site $n$. This interaction matrix element is given by%
\begin{equation}
U_{\mu^{\prime},\nu^{\prime}}^{\mu,\nu}\left(  \left\vert
n-m\right\vert \right)  =\left\langle \mu m;\nu n\left\vert V_{int}\left(
x_1-x_2\right)  \right\vert \mu^{\prime}m;\nu^{\prime}n\right\rangle ,\nonumber
\end{equation}
where $V_{int}\left(  x\right)  $ is the 1D interaction potential. In this work we will be concerned with short range interactions with
Wannier states localized to a single lattice site leading to on-site
interactions, $U_{\mu^{\prime},\nu^{\prime}}^{\mu,\nu}\left(
\left\vert n-m\right\vert \right)  =U_{\mu^{\prime},\nu^{\prime}%
}^{\mu,\nu}\delta_{m,n}$. We leave the interaction in a more general form here for completeness, and it will be specified in Sec.~\ref{subsection:twoband}.

The eigenfunctions of $\hat{H}$ can be expanded in the Wannier basis as
$\left\vert \Psi\right\rangle =\sum_{\mu\nu mn}\Psi_{ \mu,\nu}^{m,n}
\left\vert \mu m;\nu n\right\rangle $, leading to the discrete Schr\"{o}dinger
equation:%
\begin{align}
\left(  E-\epsilon_{\mu}-\epsilon_{\nu}\right)  \Psi_{ \mu,\nu}^{m,n} = & -S_{\mu}\left[  \Psi_{ \mu,\nu}^{(m+1),n}  +\Psi_{ \mu,\nu}^{(m-1),n}  \right] \label{Eq:SE1}\\
&  -S_{\nu}\left[  \Psi_{ \mu,\nu}^{m,(n+1)}  +\Psi_{ \mu,\nu}^{m,(n-1)}  \right] \nonumber\\
&  +\sum_{\mu^{\prime},\nu^{\prime}}U_{\mu^{\prime},\nu^{\prime}%
}^{\mu,\nu}\left(  \left\vert n-m\right\vert \right)  \Psi_{ \mu^{\prime},\nu^{\prime}}^{m,n} \nonumber
\end{align}

\subsection{Center of mass separation}

The most important aspect of the discrete Schr\"{o}dinger equation in Eq.~(\ref{Eq:SE1}) is that it can
be separated into the discrete center-of-mass $Z=\left(  m+n\right)  /2$ and
relative separation $z=m-n$ coordinates with the separation ansatz%
\begin{align}
\Psi_{ \mu,\nu}^{\left(  Z+z/2\right),\left(  Z-z/2\right)}
 & \propto \psi_{\mu\nu}\left(  z\right)e^{iK\lambda Z+i\phi_{K}^{\mu\nu}z}
,\label{Eq:sepansatz}
\end{align}
where $K$ is the center of mass quasi-momentum.  Here the angle
\begin{align}
\phi_{K}^{\mu\nu}  &  =\arg\left(  S_{\mu}e^{iK\lambda/2}+S_{\nu}%
e^{-iK\lambda/2}\right),
\end{align}
has been included to subtract a constant offset in the relative motion quasi-momentum.
Inserting this
ansatz into Eq.~(\ref{Eq:SE1}) yields the discrete Schr\"{o}dinger equation in the separation
coordinate,
\begin{align}
E\psi_{\mu\nu}\left(  z\right)  &=  \varepsilon_{\mu\nu}\psi_{\mu\nu}-\left[
J_{K}^{\mu\nu}\psi_{\mu\nu}\left(  z+1\right)  +J_{K}^{\mu\nu}\psi_{\mu\nu
}\left(  z-1\right)  \right] \nonumber \\
&  +\sum_{\mu^{\prime}\nu^{\prime}}e^{i\left(  \phi_{K}^{\mu^{\prime}%
\nu^{\prime}}-\phi_{K}^{\mu\nu}\right)  z}U_{\mu^{\prime}%
,\nu^{\prime}}^{\mu,\nu}\left(  \left\vert z\right\vert \right)  \psi
_{\mu^{\prime}\nu^{\prime}}\left(  z\right)  \label{Eq:SErel},
\end{align}
where the relative-coordinate hopping and two-body onsite energy are defined as
\begin{align}
J_{K}^{\mu\nu}  &=  \sqrt{S_{\mu}^{2}+S_{\nu}^{2}+2S_{\mu}S_{\mu}\cos\left(
K\lambda\right)  },\nonumber\\
\varepsilon_{\mu\nu} &=  \epsilon_{\mu}+\epsilon_{\nu}.\nonumber
\end{align}
Note that the separated Schr\"{o}dinger equation is now in the form of a simple tight-binding model with nearest-neighbor
hopping with``on-site energies" that are modified by the interaction matrix
elements $U_{\mu^{\prime},\nu^{\prime}}^{\mu,\nu}\left(  \left\vert
z\right\vert \right)  $. Also note that in the relative coordinate $z$,
the hopping energies $J_{K}^{\mu\nu}$ are now dependent on the center of
mass quasi momentum $K.$

In the absence of interactions, Eq. \ref{Eq:SErel} is solved simply by
plane waves in the relative coordinates, i.e.%
\[
\psi\left(  \mu,\nu,z\right)  \propto e^{ik\lambda z}%
\]
where $k$ is the relative coordinate quasi-momentum. The resulting dispersion
relations define two-body energy bands that depend on the center-of-mass motion  with dispersion%
\begin{equation}
\epsilon_{\mu\nu}\left(  K,k\right)  =\varepsilon_{\mu\nu}-2J_{K}^{\mu\nu}%
\cos\lambda k \label{Eq:disp}.
\end{equation}

\section{Lattice Green's function}
\label{section:latticeGF}

To examine scattering of two particles in the lattice described above, we must
first construct the Green's operator associated with the relative coordinate
Hamiltonian at fixed center of mass quasi-momentum $K$ given by%
\begin{align}
\hat{H}_{rel}  = & \hat{h}_{rel}+\hat{U}\label{Eq:Hrel}\\
\hat{h}_{rel}  = & \sum_{\mu,\nu,z}\varepsilon_{\mu\nu}\left\vert K;\mu\nu
z\right\rangle \left\langle K;\mu\nu z\right\vert \nonumber\\
&  -\sum_{\mu,\nu,z}J_{K}^{\mu\nu}\left(  \left\vert K;\mu\nu\left(
z+1\right)  \right\rangle \left\langle K;\mu\nu z\right\vert \right. \nonumber\\ 
&  + \left. \left\vert
K;\mu\nu\left(  z-1\right)  \right\rangle \left\langle K;\mu\nu z\right\vert
\right) \nonumber\\
\hat{U}  = & \sum_{\substack{\mu,\nu\\\mu^{\prime},\nu^{\prime}\\z}%
}e^{i\left(  \phi_{K}^{\mu^{\prime}\nu^{\prime}}-\phi_{K}^{\mu\nu}\right)
z}U_{\mu^{\prime},\nu^{\prime}}^{\mu\nu}\left(  \left\vert
z\right\vert \right)  \left\vert K;\mu\nu z\right\rangle \left\langle
K;\mu^{\prime}\nu^{\prime}z\right\vert \nonumber
\end{align}
where $\hat{h}_{rel}$ is the non-interacting relative Hamiltonian, and
$\hat{U}$ is the interaction. Here, we have defined the basis state
$\left\vert K;\mu\nu z\right\rangle $ as the state where the two particles
have a center of mass motion defined by quasi-momentum $K$ and a particle
separation of $z$ with particle 1 in band $\mu$ and particle 2 in band $\nu$,
i.e.
\[
\left\vert K;\mu\nu z\right\rangle =\mathcal{N}\sum_{Z=0}^{N}e^{iK\lambda
Z}\left\vert \mu\left(  Z+z/2\right)  ;\nu\left(  Z-z/2\right)  \right\rangle
\]
where $\mathcal{N}$ is a normalization constant.

We will define the lattice Green's operator $\hat{G}$ such that
\[
\left\langle K;\mu^{\prime}\nu^{\prime}z^{\prime}\left\vert \left(  \hat
{h}_{rel}-E\right)  \hat{G}\right\vert K;\mu\nu z\right\rangle =\delta_{\mu
\mu^{\prime}}\delta_{\nu\nu^{\prime}}\delta_{zz^{\prime}}.
\]
We will proceed to find $\hat{G}$ by expanding it in the Wannier states, i.e.%
\begin{equation}
\hat{G}=\sum_{\substack{\mu\nu\\z^{\prime},z}}g_{\mu\nu}\left(  K;z,z^{\prime
}\right)  \left\vert K;\mu\nu z\right\rangle \left\langle K;\mu\nu z^{\prime
}\right\vert \label{Eq:greensoperator}%
\end{equation}
where $g_{\mu\nu}\left(  K;z,z^{\prime}\right)  $ is the lattice Green's
function (LGF) which is a solution to
\begin{align}
\delta_{z,z^{\prime}}  = &  \left( \varepsilon_{\mu\nu}-E\right)  g_{\mu\nu}\left(  K;z,z^{\prime} \right)\label{Eq:latticeGFeq}\\
&-J_{K}^{\mu\nu} \left[ g_{\mu\nu}\left(  K;z+1,z^{\prime}\right) 
+ g_{\mu\nu}\left(  K;z-1,z^{\prime}\right) \right] . \nonumber
\end{align}
For the purposes of this work, we are concerned with even parity states
associated with two bosons or two spin-1/2 fermions in a singlet state.
Therefore, below we will only consider even parity solutions to
Eq.\ref{Eq:latticeGFeq}

The even parity LGF can be broken into two cases: (1) When the scattering
energy\thinspace$E$ is within the available 2-body energies $\epsilon_{\mu\nu
}\left(  K,k\right)  $ corresponding to an open two-body scattering band ; or
(2) when $E$ it is outside of the available energies defined by $\epsilon
_{\mu\nu}\left(  K,k\right)  $ corresponding to a closed two-body scattering band.

\subsection{Open band Green's function}

For energies within the $\{\mu,\nu\}$ two-body band (i.e. $\left\vert
E-\varepsilon_{\mu\nu}\right\vert <2\left\vert J_{K}^{\mu\nu}\right\vert $), we can define the relative
quasi-momentum $k$ through the dispersion relation in Eq.~(\ref{Eq:disp}).
For $z\neq z^{\prime}$ Eq. \ref{Eq:latticeGFeq} are solved by the ansatz%
\[
g_{\mu\nu}\left(  K;z,z^{\prime}\right)  =A\sin\left(  k\lambda z_{>}\right)
\cos\left(  k\lambda z_{<}\right),
\]
where $z_{>\left(  <\right)  }$ is the larger (smaller) of $z$ and $z^{\prime}%
$. Here we have chosen to use the principle value Green's function which obeys standing-wave boundary conditions. This is
similar to the approach taken in other work [(cite some others)] in which the singular portion of the Green's function corresponding to direct classical trajectories is separated and removed. The remaining constant $A$ can be found by simply inserting the ansatz into Eq. \ref{Eq:latticeGFeq} for $z=z^{\prime}$ giving
\begin{equation}
g_{\mu\nu}\left(  K;z,z^{\prime}\right)  =-\dfrac{\sin\left(  k\lambda
z_{>}\right)  \cos\left(  k\lambda z_{<}\right)  }{J_{K,}^{\mu\nu}\sin\left(
k\lambda\right)  }. \label{Eq:openbandgf}%
\end{equation}
This can be further simplified by writing it in terms of the regular, $f_{\mu\nu}^{+}$, and
irregular, $f_{\mu\nu}^{-}$, band-energy normalized scattering solutions of the noninteracting
Hamiltonian given respectively by
\begin{align}
f_{\mu\nu}^{+}\left(  z\right)   &  =\sqrt{\dfrac{\lambda}{\pi \left| v_{g}^{\mu\nu
}\right|  }}\cos\left(  k\lambda z\right)  ,\label{Eq:Enormstatea}\\
f_{\mu\nu}^{-}\left(  z\right)   &  =\sqrt{\dfrac{\lambda}{\pi \left| v_{g}^{\mu\nu
}\right|  }}\sin\left(  k\lambda z\right)  ,\nonumber\\
v_{g}^{\mu\nu}   &  =2\lambda J_{K}^{\mu\nu}\sin\left(
\lambda k\right)  =\dfrac{\partial\epsilon_{\mu\nu}\left(  K,k\right)
}{\partial k}.\nonumber
\end{align}
so that $\sum_z f^{\pm*}_{\mu\nu}(E;z)f^{\pm}_{\mu\nu}(E';z)=\delta(E-E')$.
In terms of $f_{\mu\nu}^{+}$ and $f_{\mu\nu}^{-}$ the open-band LGF is now
given by%
\begin{equation}
g_{\mu\nu}\left(  K;z,z^{\prime}\right)  =-2\pi f_{\mu\nu}^{+}\left(
z_{<}\right)  f_{\mu\nu}^{-}\left(  z_{>}\right)  \label{Eq:openLGFnormalized}%
\end{equation}

\subsection{Closed band Green's function}

For energies outside of the $\{\mu,\nu\}$ two-body energy band (when $\left\vert
E-\varepsilon_{\mu\nu}\right\vert >2\left\vert J_{K}^{\mu\nu}\right\vert $) we
have a slightly different situation, where the probability of the two particles has to vanish at large separation distances, i.e. $\left\vert z-z^{\prime}\right\vert \to \infty$. This implies that in this limit the LGF must obey exponentially decaying boundary conditions. Namely, for this case the $z\neq0$ is
solved by the ansatz \cite{Valiente2010PRA}
\[
g_{\mu\nu}\left(  K;z,z^{\prime}\right)  =A\alpha^{\left\vert z-z^{\prime
}\right\vert },
\]
where $\alpha$ is defined as the solution to%

\begin{equation}
E-\varepsilon_{\mu\nu}=-J_{K}^{\mu\nu}\dfrac{1+\alpha^{2}}{\alpha}.
\label{Eq:decaydispersion}%
\end{equation}
where we restrict $\left\vert \alpha\right\vert \leq1$. Again the coefficient
$A$ in the ansatz can be found by inserting into Eq. \ref{Eq:latticeGFeq} at
$z=z^{\prime}$ giving%
\begin{equation}
g_{\mu\nu}\left(  K;z,z^{\prime}\right)  =\dfrac{\alpha^{\left\vert
z-z^{\prime}\right\vert +1}}{J_{K}^{\mu\nu}\left(  1-\alpha^{2}\right)  }.
\label{Eq:closedbandgf}%
\end{equation}
Note that when $E<\varepsilon_{\mu\nu}-2J_{K}^{\mu\nu}$ (i.e. below the band)
Eq. \ref{Eq:decaydispersion} gives $0<\alpha<1$ such that the LGF decays
exponentially. When $E>\varepsilon_{\mu\nu}+2J_{K}^{\mu\nu}$ (i.e. above the
band) Eq. \ref{Eq:decaydispersion} gives $-1<\alpha<0$ so that the amplitude
of the LGF still decays exponentially but with alternating sign.

\section{Scattering for on-site interactions}
\label{section:scattering}
Here we will use the Green's operator found above to extract scattering
properties of two particles in a 1D lattice interacting via on-site
interactions only. The full Schr\"{o}dinger equation for the relative motion can be solved via the Lippmann-Schwinger
equation (LSE) given by%
\begin{equation}
\left\vert \psi\right\rangle =\left\vert \psi_{0}\right\rangle -\hat{G}\hat
{U}\left\vert \psi\right\rangle ,
\end{equation}
where the homogeneous solution $\left\vert \psi_{0}\right\rangle =\sum_{z}%
f_{\mu\nu}^{+}\left(  z\right)  \left\vert K;\mu\nu z\right\rangle $ is the
initial state of the system. Note here that we are using the band-energy normalized
scattering states from Eq. \ref{Eq:Enormstatea}. In the $z\rightarrow\infty$
limit, inserting the LGF from Eq. \ref{Eq:openLGFnormalized} gives the
scattering solution%
\begin{align}
\psi_{\mu\nu}\left(  z\right)   = & f_{\mu\nu}^{+}\left(  z\right)  +2\pi
f_{\mu\nu}^{-}\left(  z\right)  \left\langle f_{\mu\nu}^{+}\left\vert \hat
{U}\right\vert \psi\right\rangle ,\label{Eq:scatterinsols}\\
\left\langle f_{\mu\nu}^{+}\left\vert \hat{U}\right\vert \psi\right\rangle  
= & \sum_{\mu^{\prime}\nu^{\prime}z^{\prime}}f_{\mu\nu}^{+}\left(  z^{\prime
}\right)  e^{i\left(  \phi_{K}^{\mu^{\prime}\nu^{\prime}}-\phi_{K}^{\mu\nu
}\right)  z^{\prime}} \nonumber\\
& \times U_{\mu^{\prime},\nu^{\prime}}^{\mu\nu}\left(
\left\vert z^{\prime}\right\vert \right)  \psi_{\mu^{\prime}\nu^{\prime}%
}\left(  z^{\prime}\right). \nonumber
\end{align}
Here we assume the interaction
between the two particles is short range in comparison with lattice's periodicity, i.e. $\ell_{int}\ll\lambda$ where $\ell_{int}$ is
the length scale of the inter-particle interaction. Additionally, we assume that the Wannier states are localized to single lattice sites. These two
assumptions, sufficiently, imply that the particles interact via on-site interactions only,
$U_{\mu^{\prime}\nu^{\prime}}^{\mu\nu}\left(  \left\vert
z\right\vert \right)  =U_{\mathbf{ij}}\delta_{z,0}$, where the double band index $\{\mu,\nu\}$ is  collectively denoted by a single index vector, i.e.
$\mathbf{i}=\left\{  \mu,\nu\right\}  $, $\mathbf{j}=\left\{  \mu^{\prime}%
,\nu^{\prime}\right\}  $, and the onsite interaction matrix elements are given
by $U_{\mathbf{ij}}=U_{\mu^{\prime}\nu^{\prime}}^{\mu\nu}$.

In addition, Eq.~\ref{Eq:scatterinsols} can be generalized to include transitions
between two open bands with overlapping energies. From this the lattice
K-matrix element $\mathbf{K}_{\mathbf{ji}}^{L}$ is identified which in return determines the admixture
of the irregular solution $f^{(-)}_{\mathbf{j}}\left(  z\right)  $ in the final band
$\mathbf{j}$:%
\begin{align}
\left\langle K;\mathbf{j}z|\psi\right\rangle  &  =\delta_{\mathbf{ji}%
}f_{\mathbf{i}}^{+}\left(  z\right)  -K_{\mathbf{ji}}^{L}f_{\mathbf{j}}%
^{-}\left(  z\right)  ,\label{Eq:Kmatgen}\\
\mathbf{K}_{\mathbf{ji}}^{L}  &  =-2\pi\left\langle f_{\mathbf{j}}%
^{+}\left\vert \hat{U}\right\vert \psi\right\rangle ,\nonumber
\end{align}
where $\left\vert f_{\mathbf{j}}^{+}\right\rangle =\sum_{z}f_{\mathbf{j}}%
^{+}\left(  z\right)  \left\vert \mathbf{K};\mathbf{j}z\right\rangle .$ Solving this
equation self-consistently for the $K$-matrix yields,%
\begin{equation}
\mathbf{K}_{\mathbf{ji}}^{L}=-2\pi\left\langle f_{\mathbf{j}}^{+}\left\vert
\left(  \mathbb{1}+\hat{U}\hat{G}\right)  ^{-1}\hat{U}\right\vert f_{\mathbf{i}%
}^{+}\right\rangle , \label{Eq:Kmatselfconsist}%
\end{equation}
where \thinspace$\hat{G}$ and $\hat{U}$
indicate the LGF and interaction operators, respectively. 

In the case of on-site interactions, $\hat{U}=\sum_{\mathbf{ij}}%
U_{\mathbf{ij}}\left\vert K;\mathbf{i}0\right\rangle \left\langle
K;\mathbf{j}0\right\vert $, we show in Appendix A that by partitioning
$\hat{G}$ and $\hat{U}$ into $N$ open and $M$ closed band contributions the
lattice $K$-matrix for scattering from one open band to another is given by%
\begin{equation}
\mathbf{K}^{L}=-2\lambda \bar{v}_{g}^{-1/2}\left[
\mathcal{U}_{oo}-\mathcal{U}_{oc}\bar{g}_{c}\left(  \mathbb{1}+\mathcal{U}_{cc}\bar
{g}_{c}\right)  ^{-1}\mathcal{U}_{co}\right]  \bar{v}_{g}^{-1/2}. \label{Eq:Kmatfinal}%
\end{equation}
Here $\mathcal{U}_{oo}$ is the $N\times N$ matrix of interaction matrix
elements $U_{\bf ij}$ for initially and finally open bands.
$\mathcal{U}_{co}=\left[  \mathcal{U}_{oc}\right]  ^{\dag}$ is the $M\times N$
matrix of interaction matrix element between the finally closed and initially open 
bands. $\mathcal{U}_{cc}$ is the $M\times M$ matrix of interaction matrix
elements for initially and finally closed bands. The $M\times M$ diagonal
matrix $\bar{g}_{c}$ has a diagonal of closed channel LGFs evaluated at
$z=z^{\prime}=0$. Finally, $\left[  \bar{v}_{g}\right]  ^{-1/2}$ is an
$N\times N$ diagonal matrix with diagonal values given by the inverse of the
square root of the open-band group velocities$\left(  v_{g}^{\mathbf{i}%
}\left(  k_{\mathbf{i}}\right)  \right)  ^{-1/2}.$

The form of Eq. \ref{Eq:Kmatfinal} is familiar in scattering theory, being
quite reminiscent of the standard channel closing formulas of multi-channel
quantum defect theory \cite{aymar1996mrs}. The first term describes the
background scattering in the open bands. The second term incorporates the contributions from virtual scattering into
energetically closed two-body bands. Notice that including these closed band
terms allows for resonances when $\det\left(   \mathbb{1}+\mathcal{U}_{cc}\bar
{g}\right)  =0$ in the open band K-matrix $\mathbf{K}^{L}$. 
In essence, this means that all these virtual transitions in closed bands can collectively give rise to lattice induced resonances in the open bands.
Also note that the K-matrix can related to the standard S-matrix via the expression  $\mathbf{S}%
^{L}\mathbf{=}\left(  \mathbb{1}+i\mathbf{K}^{L}\right)  \left(   \mathbb{1}-i\mathbf{K}%
^{L}\right)  ^{-1}$.

In the absence of other bands, Eq.~\ref{Eq:Kmatfinal} gives
that the K-matrix for single band scattering with onsite interactions is
simply proportional to the interaction strength as expected.
By analogy to this, in a single open band, the contributions from excited
bands results in a quasi-momentum dependent effective interaction given by%
\begin{equation}
U_{eff}=\mathcal{U}_{oo}-\mathcal{U}_{oc}\bar{g}_{c}\left(   \mathbb{1}+\mathcal{U}%
_{cc}\bar{g}_{c}\right)  ^{-1}\mathcal{U}_{co}. \label{Eq:Ueff}%
\end{equation}
Properly including these effects in many-body models like the Bose-Hubbard
model is likely quite important especailly in the presenece of the above
mentiond lattice induced resonances.

\subsection{Two-band approximation}
\label{subsection:twoband}

Here we simplify our system by assuming a simple two-band approximation. We
will assume that each particle is in the Wannier states $w_{\mu}\left(
x\right)  $ corresponding to either the lowest band ($\mu=0$) or the first
excited band ($\mu=1$) meaning that the available two-body states are
restricted to $\left\{  \mu,\nu\right\}  =\left\{  0,0\right\}  ,\left\{
0,1\right\}  ,$ $\left\{  1,0\right\}  ,$ and $\left\{  1,1\right\}  $. If we
assume that the interaction potential is symmetric under inversion and the
Wannier states are parity eigenstates, then the two-body state with one
particle in the excited band is decoupled by parity, i.e. $U_{00}^{10}%
=U_{11}^{10}=0$. We will further assume that the interaction potential
$V\left(  x\right)  $ is short range enough to be approximated by a contact
interaction. For notation simplicity we will label the interaction matrix
elements $U_{\mu\nu}^{\mu^{\prime}\nu^{\prime}}$ as $U_{00}^{00}=U_{00},$
$U_{01}=U_{10}=U_{00}^{11}=U_{11}^{00};$ and $U_{11}=U_{11}^{11}$.

If both particles start in the lowest $\left\{  0,0\right\}  $ band such that
the $\left\{  1,1\right\}$ band is energetically inaccessible, Eq.
(\ref{Eq:Kmatfinal}) becomes%
\begin{align}
\mathbf{K}_{00\rightarrow00}^{L}= & \dfrac{-1}{J_{K}^{00}\sin\lambda k}\label{Eq:2BandScattering1}\\
& \times \left[U_{00} - \dfrac{U_{01}^{2}}{U_{11}+\sqrt{\left(  E-\varepsilon_{11}\right)
^{2}-\left(  2J_{K}^{11}\right)  ^{2}}}\right] \nonumber. %
\end{align}
In the case where the onsite interactions in the excited band are attractive,
i.e. $U_{11}<0$, a resonance occurs at precisely the energy of a dimer bound
state attached to the \emph{excited} band, $E_{dim}=\varepsilon_{11}-\sqrt
{U_{11}^{2}+\left(  2J_{K}^{11}\right)  ^{2}}$. Thus, we see that a lattice
induced resonance occurs due to virtual scattering into a bound state attached
to an excited band that energetically lies in the continuum of the lower band. Intuitively, this means that the lattice induced resonances fulfill a Fano-Feshbach-like scenario where the continuum is structured into bands due to the presence of the lattice.

Unlike scattering in free space, the energy bands induced by the lattice allow
the existence of scattering channels at energies \emph{below} the scattering energy
that are energetically inaccessible. In the case of the two-band approximation
this means that two-particles scattering in the excited band can go through
virtual scattering processes in a lower, energetically inaccessible, band. 
The K-matrix for the
$\left\{  1,1\right\}  \rightarrow\left\{  1,1\right\}  $ scattering process
is given from Eq. \ref{Eq:Kmatfinal} as
\begin{align}
\mathbf{K}_{11\rightarrow11}^{L}= & \dfrac{-1}{J_{K}^{11}\sin\lambda k} \label{Eq:2BandSacttering2} \\
& \times \left[
U_{11}-\dfrac{U_{01}^{2}}{U_{00}-\sqrt{\left(  E-\varepsilon_{00}\right)
^{2}-\left(  2J_{K}^{00}\right)  ^{2}}}\right]  . \nonumber
\end{align}
In this case we can see that a scattering resonance occurs for $U_{00}>0$
when a state is bound \emph{above} the lowest band at energy $E_{dim}=\varepsilon
_{00}+\sqrt{U_{00}^{2}+\left(  2J_{K}^{00}\right)  ^{2}}$ is embedded in the
excited two-body band. Counter-intuitively, this brings about the possibility for \emph{repulsive} on-site interactions in an energetically closed lower band inducing
\emph{resonant} interactions in an excited band. This might be relevant for
the case of spin 1/2 fermions in which the Fermi level is at the bottom of an
excited band. In this case, it might be possible for repulsive onsite
interactions between opposite spin particles in the lowest band to induce
strong effectively attractive interactions in particles at the Fermi level in the conduction band.

\begin{figure}[ht!]
\includegraphics[width=3in]{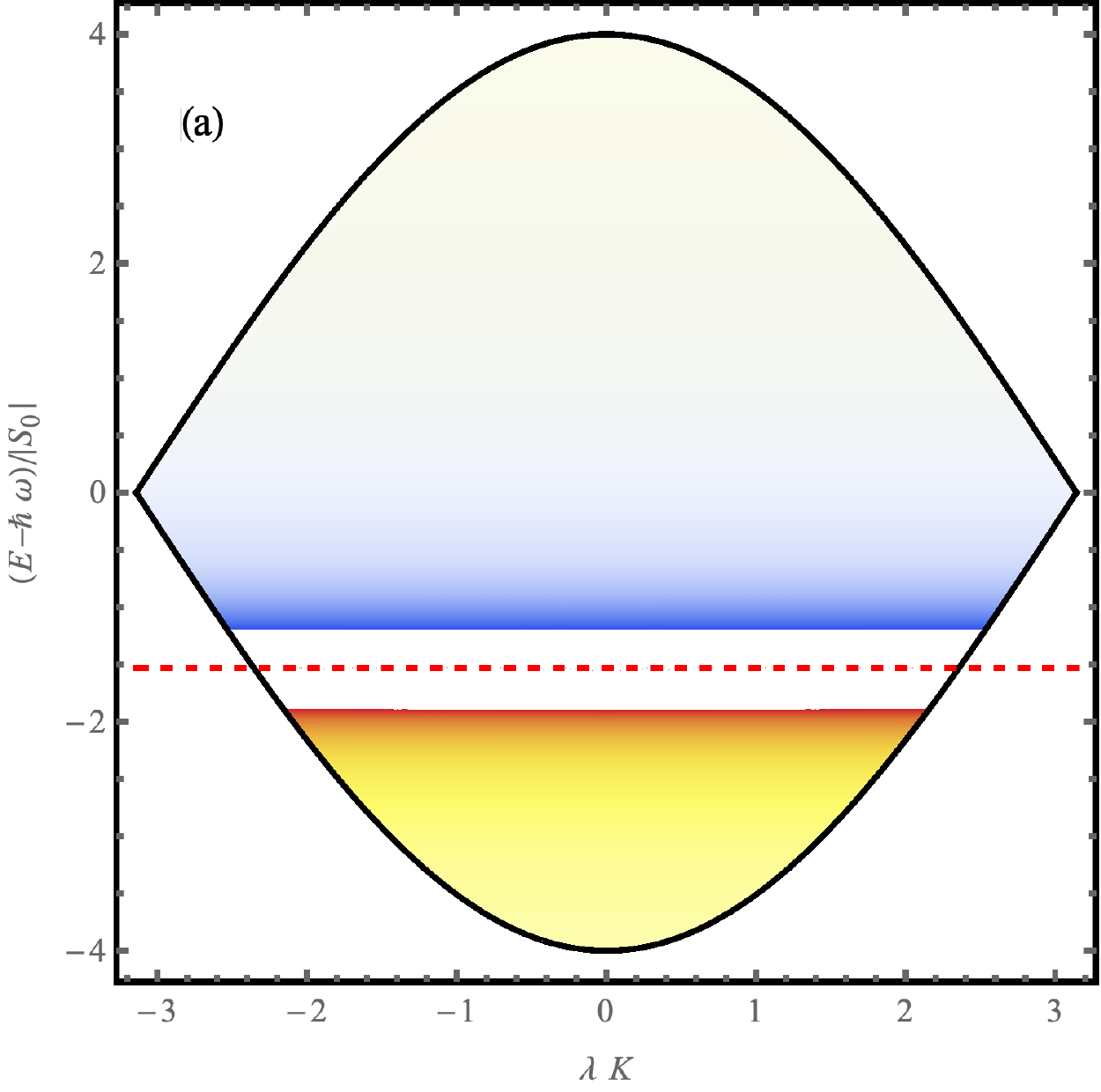}
\includegraphics[width=3in]{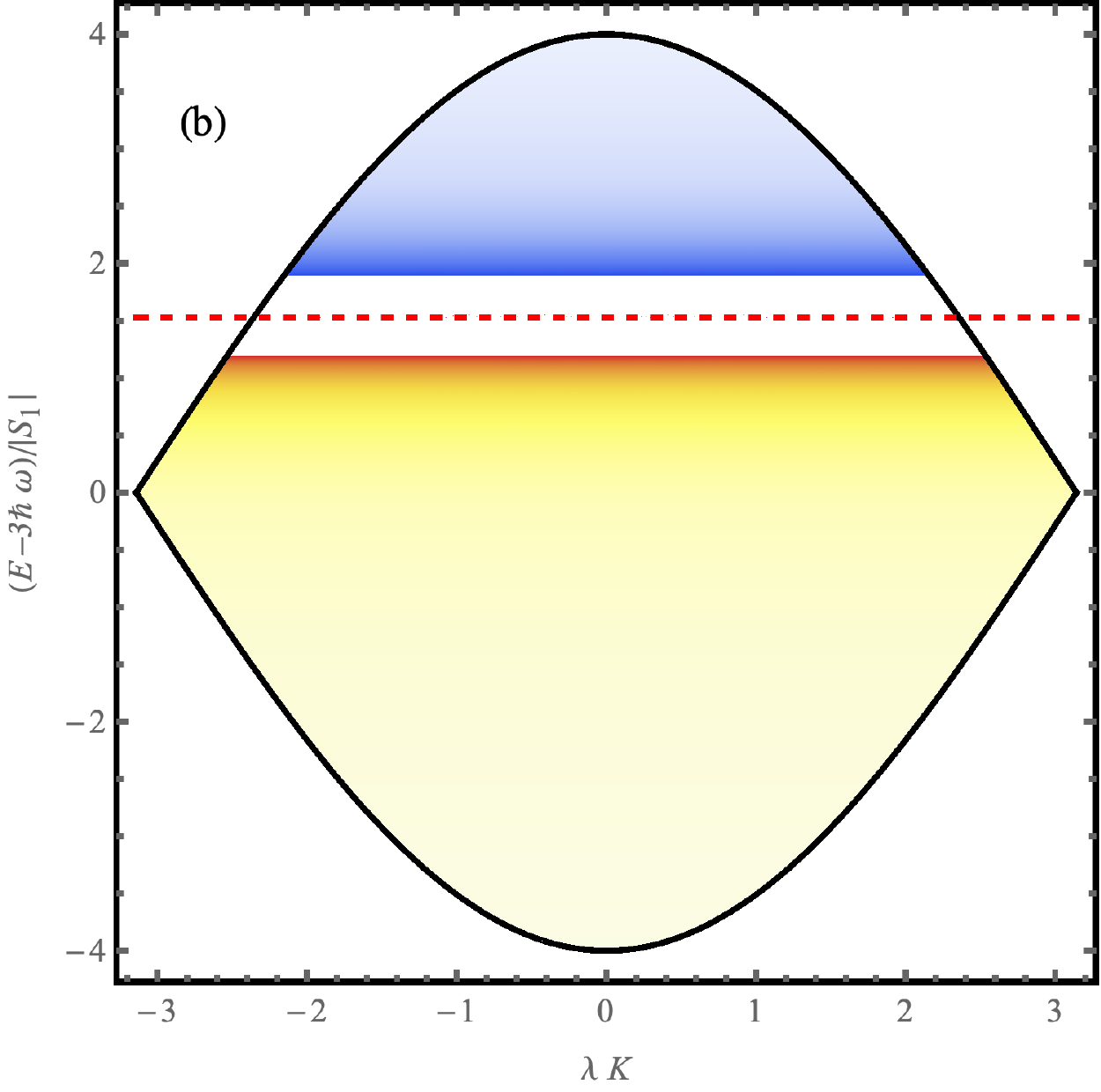}\caption{(color online) $J_{K}^{\mu\nu
}\sin\left(  k\lambda\right)  \mathbf{K}_{\mu\nu\rightarrow\mu\nu}^{L}$ under the
two-band approximation with harmonic oscillator Wannier states is shown for
(a) scattering in the lowest 2-body band $\{0,0\}\rightarrow\{0,0\}$ with
$\lambda=8\ell_{HO}$ and $a_{1D}$ 
set so that the bound excited band bound state intersects the upper two-body bad at $\lambda K = 3\pi/4$. The same is  shown in (b) for scattering in the first
excited band $\{1,1\}\rightarrow\{1,1\}$ with $\lambda=8\ell_{HO}$ and
$a_{1D}$ 
 set so that the bound state attached to the lower band intersects the lower portion of the two-body band at $\lambda K = 3\pi/4$. The lower yellow region and upper blue regions
represent positive and negative values of K respectively, with darker color
representing a larger magnitude. In both (a) and (b) the solid black curves
mark the edge of the respective 2-body bands while the dashed red curves
represent the position of a bound state attached to the (a) excited and (b)
ground 2-body bands.}%
\label{Fig:KMatHO}%
\end{figure}

Figure \ref{Fig:KMatHO} shows the K-matrix for scattering in the lowest and
first excited two-body bands assuming that the lattice potential is deep
enough such that the lowest several Wannier states can be approximated by
oscillator states, i.e. $w_{\mu}\left(  x\right)  \propto e^{-x^{2}/2\ell_{HO}^{2}%
}H_{n}\left(  \dfrac{x}{\ell_{HO}}\right)  $, where $\ell_{HO}$ is the local osciallator
length near the bottom of a lattice site and $H_{n}\left(  y\right)  $ is a
Hermite polynomial. Here, the interaction potential is taken to be a simple
contact interaction whose strength is governed by the 1D free space scattering
length:
\[
V_{int}\left(  x_1-x_2\right)  =-\dfrac{2\hbar^{2}}{ma_{1D}}\delta\left(  x_1-x_2\right)
.
\]
Experimentally, in the presence of a strong transverse confinement, $a_{1D}$ could be tuned using a confinement induced resonance \cite{bergeman_etal}.
With these assumptions we can calculate all of the relevant parameters from
Eqs. \ref{Eq:onebodham} and \ref{Eq:fullham}. The resulting K-matrix for
scattering in the lowest band, $\mathbf{K}_{00\rightarrow00}^{L}$ is shown in Fig. \ref{Fig:KMatHO}(a) plotted as a function
of $K$ in units of $1/\lambda$ and $E$ shifted to the center of the \{0,0\} band in units of $\left| S_0 \right|$.  The local
oscillator length is set so that $\ell_{HO}=\lambda/8$. The lattice-free  1D scattering length
$a_{1D}$ has been set by requiring the binding energy of a bound state attached to the excited band to intersect the upper part of the \{0,0\} two-body band at center-of-mass quasi-momentum $\lambda K = 3\pi/4$. 
Also shown is the
energy of the bound state attached to the excited 2-body band, indicated by the red dashed line. Here we can
clearly see the resonance that occurs when the bound state is at energies
accessible in the band. Figure \ref{Fig:KMatHO}(b) shows the K-matrix for
scattering in the excited band, $\mathbf{K}_{11\rightarrow11}^{L}$, for the
same lattice parameters with a 1D lattice-free scattering length set to be negative with a bound state attached to the \emph{lower} band intersecting the lower portion of the \{1,1\} two-body band at $\lambda K = 3\pi/4$. Note that in both Fig. \ref{Fig:KMatHO}(a) and (b), we have
multiplied the K-matrix by $J_{K}^{\mu\nu}\left(  K\right)  \sin\left(
\lambda k\right)  $ to remove the singularities at the edge of the two-body
band (when $\lambda k=0,\pm\pi$). The energy of the bound state attached to
the lower two-body band is shown as a dashed red line in Fig. \ref{Fig:KMatHO}%
b. As the state cuts through the band, the related a scattering resonance can
be seen in $\mathbf{K}_{11\rightarrow11}^{L}$.

\subsection{Beyond two band approximation}

In the case of scattering in the lowest two-body band in the presence of more
than one excited band, if the excited bands are uncoupled, the above results
can be easily extended to give%
\begin{align*}
\mathbf{K}_{00\rightarrow00}^{L}=&-\dfrac{1}{J_{K}^{00}\sin\lambda k}\left[
U_{00}^{00}\right.\\
 - & \left.\sum_{\left\{  \mu\nu\right\}  \neq\left\{  00\right\}  }%
\dfrac{\left\vert U_{\mu\nu}^{00}\right\vert ^{2}}{U_{\mu\nu}^{\mu\nu}%
+ \sqrt{\left(  E-\varepsilon_{\mu\nu}\right)  ^{2}-\left(  2J_{K}^{\mu\nu
}\right)  ^{2}}}\right]  .
\end{align*}
The sum here accounts for virtual scattering into each excited band. Notice
that a properly tuned attractive interaction diagonal matrix element in an
excited band, $U_{\mu\nu}^{\mu\nu}$, can create a bound state attached to that
two-body band that cuts through the lowest band creating a scattering
resonance, similarly to Fig.\ref{Fig:KMatHO} (a).
Coupling between excited bands can shift the position of
these bound states 
However, even with these shifts, if the states become
degenerate with the $\left\{  \mu\nu\right\}  =\left\{  00\right\}  $ band, we
expect a band-induced scattering resonances to occur.

In the case of scattering in excited bands, it is possible for multiple
excited two-body bands to overlap in energy. In this case, whenever the scattering energy
and center of mass quasi-momentum place the system in the overlap region of multiple two-body bands,
Eq.~\ref{Eq:Kmatfinal} still hold, but the K-matrix is an $N\times N$ matrix where $N$ is the number of overlapping bands.  The diagonal elements of $\mathbf{K}$ represent elastic scattering processes where the incident and outgoing states are in the same bands.  However, the off diagonal elements represent inelastic scattering processes where the energy and center-of-mass quasi-momentum of the system $K$ is conserved, but the relative quasi-momentum $k$ is not.

\begin{figure*}
\includegraphics[width=6in]{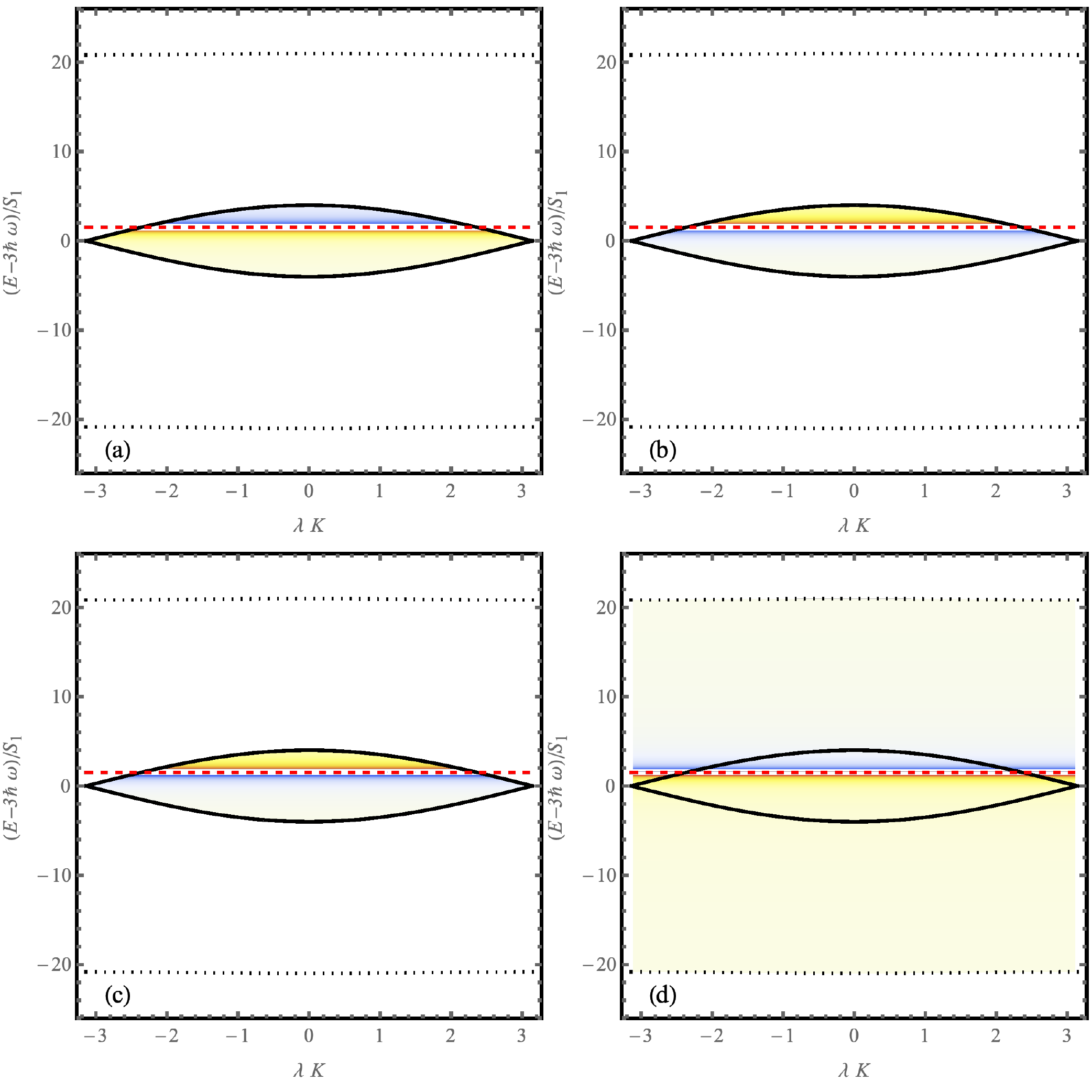}
\caption{The  matrix elements of $\bar{v}_{g}^{1/2}\mathbf{K}^L\bar{v}_{g}^{1/2}$
for scattering are shown as a density plots with harmonic oscillator Wannier states.  Here we have included the $\{0,0\}$, $\{1,1\}$, 
and the symmetrized $\{0,2\}$ two-body bands, and are showing the K-matrix for scattering in the overlapping excited states. (a) and (d) show the matrix elements for elastic scattering in the $\{1,1\}$ and $\{0,2\}$ states respectively.  (b) and (c) show the matrix elements for inelastic scattering $\{1,1\}\rightarrow\{0,2\}$ and $\{0,2\}\rightarrow\{1,1\}$ respectively.  The lattice and interaction parameters are the same as in Fig.~\ref{Fig:KMatHO}(b).}
\label{Fig:K2x2MatHO}%
\end{figure*}

For example, in the case where
the lattice sites are deep enough to be treated locally as harmonic
oscillators, the $\left\{  \mu\nu\right\}  =\left\{  11\right\}  $ band
overlaps with the $\left\{  20\right\}  $ and $\left\{  02\right\}  $ bands
meaning that the lattice K-matrix, $\mathbf{K}^L$, is a $3\times3$ matrix (or $2\times2$ in the case of symmetrized states for bosonic scattering). The K-matrix elements are shown in Fig.~\ref{Fig:K2x2MatHO}(a-d) for the case of the $\{1,1\}$ and the symmetrized $\{0,2\}$ two-body bands overlapping .  Here, the lattice and interaction parameters are set to be the same as in Fig.~\ref{Fig:KMatHO}(b).The solid and dotted black lines show the edges of the $\{1,1\}$ and the symmetrized $\{0,2\}$ two-body bands
respectively.  The position of a resonant state attached to the lower $\{0,0\}$ band is shown as the red dashed curve.  The $\{0,2\}$ band complete encloses the $\{1,1\}$ band.  Figures \ref{Fig:K2x2MatHO}(b) and (c) showing the $\mathbf{K}_{11\rightarrow02}$ and $\mathbf{K}_{02\rightarrow11}$ inelastic K-matrix element respectively are identical.  According to Eq.~\ref{Eq:Kmatfinal} the $\mathbf{K}_{11\rightarrow11}$ matrix element is the same as in the two-band case given in Eq.~\ref{Eq:2BandSacttering2} and thus Fig.~\ref{Fig:K2x2MatHO}(a) is the same as Fig.~\ref{Fig:KMatHO}(b) but plotted on a different energy scale. Notice that a resonance appears at the same energy for all K-matrix elements. 

Notice in Eq.~\ref{Eq:Kmatfinal} that the resonance condition $\det\left(  \mathbb{1}+\mathcal{U}_{cc}\bar
{g}\right)  =0$ deals only with information from the closed bands.  Thus if a resonance appears in the diagonal K-matrix elements (the elastic scattering processes), it will appear at the same energy in the off diagonal elements (in the inelastic scattering processes).

\section{The Lattice Scattering Length}
\label{section:scatlen}

Just as in normal lattice-free scattering in 1D, we can define the 1D lattice
scattering length. In contrast, the scattering length in the presence of a lattice
can be defined as the relative quasi-momentum reaches the edge of a two-body
band either at the top or the bottom of the band, here when $\lambda k\rightarrow0,\pi$:%
\begin{align}
\lim_{k\rightarrow0}k\mathbf{K}_{\mu\nu\rightarrow\mu\nu}^{L}  &  =\dfrac{1}{a_{\mu\nu
}^{\left(  -\right)  }},\label{Eq:scattlength0}\\
\lim_{k\rightarrow\pi/\lambda}\left(  \dfrac{\pi}{\lambda}-k\right) \mathbf{K}_{\mu
\nu\rightarrow\mu\nu}^{L}  &  =\dfrac{1}{a_{\mu\nu}^{\left(  +\right)  }%
}.\nonumber
\end{align}
Here $a_{\mu\nu}^{\left(  -\right)  }$ and $a_{\mu\nu}^{\left(  +\right)  }$
is the scattering length at the bottom and top of the two-body bad
respectively corresponding to elastic scattering in the $\left\{  \mu
\nu\right\}  $ band. In the two-band approximation from above this yields%
\begin{align}
\dfrac{\lambda J_{K}^{00}}{a_{00}^{\left(  \pm\right)  }}  &  =-\left[  U_{00}-\dfrac{U_{01}^{2}}{U_{11}+\sqrt{\left( E^{(\pm)}_{00} -\varepsilon_{11}\right)  ^{2}-\left(  2J_{K}^{11}\right)
^{2}}}\right]  \label{Eq:ScattLengt00},\\
\dfrac{\lambda J_{K}^{11}}{a_{11}^{\left(  \pm\right)  }}  &  =-\left[  U_{11}-\dfrac{U_{01}^{2}}{U_{00}-\sqrt{\left(E^{(\pm)}_{11}-\varepsilon_{00}\right)  ^{2}-\left(  2J_{K}^{00}\right)
^{2}}}\right] \label{Eq:ScattLengt11} .
\end{align}
Here $E^{(\pm)}_{\mu\nu}=\varepsilon_{\mu\nu}\pm2J_{K}^{\mu\nu}$ is the energy at the top ($+$) and bottom ($-$) of the band.

\begin{figure}[ht!]
\includegraphics[width=3in]{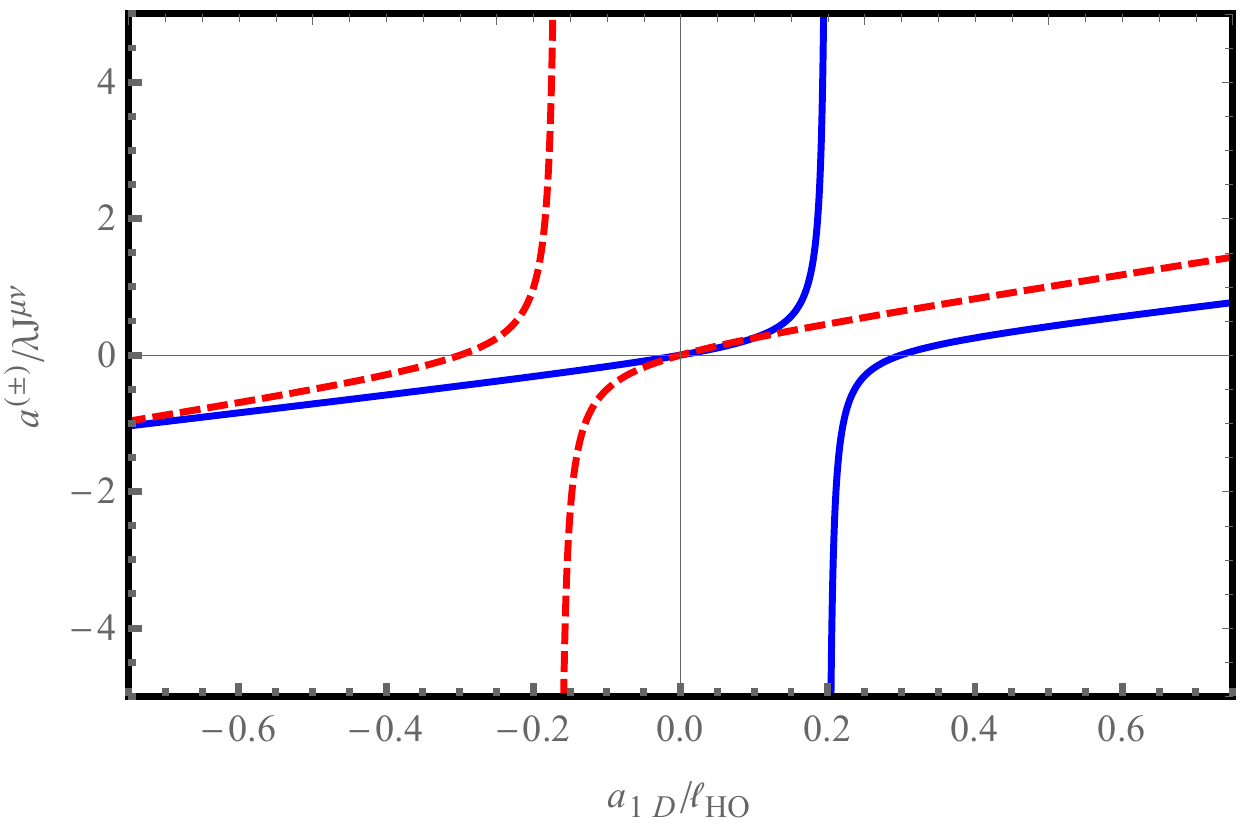} \caption{(color online) The
lattice scattering length $a^{\left( \pm \right)  }$ is shown in the $\{0,0\}$ (blue solid curve) and $\{1,1\}$ (red dashed curve) two-body bands is shown as a function of the 1D scattering length $a_{1D}$ for a lattice with lattice spacing $\lambda= 8\ell_{HO}$.  For a lattice this deep, the hopping energy is much smaller than the local oscillator energy and thus $a^{(+)}\approx a^{(-)}$.}%
\label{Fig:scatlength}%
\end{figure}

Contrary to the 3D case, strong effective interactions occur in 1D near zeros
in the scattering length. Conversely, poles in the scattering length occur at
zeros in the lattice K-matrix near the top or bottom of the two-body bands.
Figure \ref{Fig:scatlength} shows the lattice scattering length $a_{\mu\nu
}^{\left(  \pm\right)  }$ for the lowest and first excited two-body band
within the two-band approximation
plotted as a function of the free-space 1D scattering length $a_{1D}$. We have
again assumed that the lattice sites are deep enough to be treated as local
harmonic oscillator with lattice spacing $\lambda=8\ell_{HO}$. We can clearly see
that at finite values of the 1D free-space scattering length, there are poles
in the lattice scattering length corresponding to areas of weak effective interaction.
The zeros in the lattice scattering length correspond to strong, resonant
effective interactions. While we show the $K=0$ scattering lengths here,
similar structures with small shifts appear for all values of the
center-of-mass quasi-momentum. In the case of a deep lattice such as that used here, the hopping energy becomes much smaller than the local oscillator energy, and thus much smaller than the band gap energy.  When the on-site interaction energy is much larger than the hopping energy for all center of mass quasi-momenta ($\left| U_{\mu\nu}^{\mu'\nu'} \right| \gg \left| J^{\mu\nu}\right|$) the quasi-momentum dependence of the lattice scattering length drops out from the right hand side of Eqs.~(\ref{Eq:ScattLengt00}) and (\ref{Eq:ScattLengt11}). In addition, when the hopping energy is small compared to the interaction energy, there is effectively no difference between $a^{(+)}_{\mu\nu}$ and $a^{(-)}_{\mu\nu}$.

Large positive lattice scattering lengths correspond to a weakly bound state.
Within the lattice approximations made here, the energy of the bound state is
given approximately by
\begin{equation}
E_{bnd}\approx\varepsilon_{\mu\nu}\pm J_{K}^{\mu\nu}\sqrt{4+\left(\dfrac{\lambda}{
a_{\mu\nu}^{\left(  \pm\right)  }}\right)^2},\label{Eq:Ebnd}%
\end{equation}
with the approximation becoming exact at unitarity, i.e. $\left|a_{\mu\nu}^{\left(  \pm\right)
}\right|\rightarrow\infty$. Note that poles in the lattice scattering length occur when that
bound state becomes degenerate with the two-body band continuum.

\section{Summary}
\label{section:summary}

In this study we explored two-body scattering in the presence of a one dimensional lattice. By transforming into a basis of Wannier states and removing the discrete center-of-mass position we derived the multi-band Green's operator using a Lattice Green's function for either energetically open or closed bands. This Green's operator was then used with in the Lippmann-Schwinger equation to extract the lattice K-matrix. 

In the case of on-site interactions, the K-matrix consists of two terms, the first, which is proportional to the open-channel interaction matrix, correspond to scattering between the energetically open two-body bands, while the second term accounts for virtual scattering events into energetically closed bands allowing for resonant scattering.  In the absence of coupling between closed band, resonances occur when bound states attached to closed bands are embedded in the open bands.

The expression for the scattering K-matrix derived here incorporates the
scattering contributions from any number of overlapping open bands with
any number of closed bands. In deriving Eq. \ref{Eq:Kmatfinal} we have
assumed nearest neighbor hopping and on-site interactions only. However, as the band index increases,
the contributions from distant hopping will become larger, and will not
necessarily be negligible. Additionally, higher index Wannier states will
become less and less localized to the point where individual states span
multiple sites creating interactions beyond the onsite ones making the contact potential approximation invalid. Higher band contributions in 3D lattices were directly incorporated for
scattering in the lowest band in the zero center-of-mass quasi-momentum regime
in Refs.~\cite{fedichevExtendedMoleculesGeometric2004} and \cite{Cui2010PRL}. Properly incorporating higher energy bands in the K-matrix as well as extending these results to higher dimensions is the focus of
ongoing work.  

\section*{Apendix A}

Starting from the K-matrix element given in Eq. \ref{Eq:Kmatselfconsist} we
wish to show the result of Eq. \ref{Eq:Kmatfinal} in the case of on-site
interactions. Expanding the scattering states $\left\vert f_{\mathbf{i}\left(
\mathbf{j}\right)  }\right\rangle $ and inserting a complete set of Wannier
states yields%
\begin{align}
\mathbf{K}_{\mathbf{ji}}^{L}  &  =-2\pi\left[  \sum_{\mathbf{l\in}\text{open}%
}\sum_{z,z^{\prime}}f_{\mathbf{j}}^{+}\left(  z^{\prime}\right)  \left\langle
\mathbf{j}z^{\prime}\left\vert \hat{D}^{-1}\right\vert \mathbf{l}%
z\right\rangle \left\langle \mathbf{l}z\left\vert \hat{U}\right\vert
\mathbf{i}z\right\rangle f_{\mathbf{i}}^{+}\left(  z\right)  \right.
\label{Eq:AppKmat}\\
&  +\left.  \sum_{\mathbf{l\in}\text{closed}}\sum_{z,z^{\prime}}f_{\mathbf{j}%
}^{+}\left(  z^{\prime}\right)  \left\langle \mathbf{j}z^{\prime}\right\vert
\hat{D}^{-1}\left\vert \mathbf{l}z\right\rangle \left\langle \mathbf{l}%
z\left\vert \hat{U}\right\vert \mathbf{i}z\right\rangle f_{\mathbf{i}}%
^{+}\left(  z\right)  \right] \nonumber
\end{align}
where $\hat{D}=\left(  \mathbb{1}+\hat{U}\hat{G}\right)  $ and we have dropped
the center of mass quasi-momentum $K$ dependence everywhere for notational
simplicity. Note that we have split the sum over the band indices into the
contributions from the $N$ open and $M$ closed bands. We have used the fact
that $\left\langle \mathbf{l}z^{\prime\prime}\right\vert \hat{U}\left\vert
\mathbf{i}z\right\rangle =\left\langle \mathbf{l}z\right\vert \hat
{U}\left\vert \mathbf{i}z\right\rangle \delta_{zz^{\prime\prime}}$. We now
wish to invert $\hat{D}$ in the basis of Wannier states which can be broken
into 4 blocks:%
\[
\bar{D}=\left(
\begin{array}
[c]{cc}%
\bar{D}_{oo} & \bar{D}_{oc}\\
\bar{D}_{co} & \bar{D}_{cc}%
\end{array}
\right)
\]
where $\bar{D}$ is the operator $\hat{D}$ expressed as a matrix in the Wannier
basis whose the matrix elements are given by%
\begin{align*}
\bar{D}_{\mathbf{jl}}\left(  z^{\prime},z\right)  =  &  \left\langle
\mathbf{j}z^{\prime}\left\vert \hat{D}\right\vert \mathbf{l}z\right\rangle \\
=  &  \delta_{\mathbf{jl}}\delta_{z^{\prime}z}\\
&  +e^{i\left(  \phi_{K}^{\mathbf{l}}-\phi_{K}^{\mathbf{j}}\right)  z^{\prime
}}U_{\mathbf{jl}}\left(  \left\vert z^{\prime}\right\vert \right)
g_{\mathbf{l}}\left(  z,z^{\prime}\right)  .
\end{align*}
Here $\bar{D}_{oo}\left(  z,z^{\prime}\right)  $ is an $N\times N$ matrix
where both $\mathbf{j}$ and $\mathbf{l}$ correspond to open bands, $\bar
{D}_{co}\left(  z,z^{\prime}\right)  $ is an $M\times N$ matrix where
$\mathbf{j}$ is a closed band and $\mathbf{l}$ is open, $\bar{D}_{oc}\left(
z,z^{\prime}\right)  $ is an $N\times M$ matrix where $\mathbf{j}$ is an open
band and $\mathbf{l}$ is closed, and $\bar{D}_{cc}\left(  z,z^{\prime}\right)
$ is an $M\times M$ matrix where both $\mathbf{j}$ and $\mathbf{l}$ correspond
to closed bands. Notice that the only difference between closed and open bands
is the LGF $g_{\mathbf{l}}\left(  z,z^{\prime}\right)  $ used. Thus the form
of the matrix elements for $\bar{D}_{oo}\left(  z,z^{\prime}\right)  $ and
$\bar{D}_{co}\left(  z,z^{\prime}\right)  $ are the same given by%
\begin{align}
\bar{D}_{\mathbf{jl}}\left(  z^{\prime},z\right)  =  &  \delta_{\mathbf{jl}%
}\delta_{z^{\prime}z}\label{Eq:Doomatelem}\\
&  +2\pi e^{i\left(  \phi_{K}^{\mathbf{l}}-\phi_{K}^{\mathbf{j}}\right)
z^{\prime}}U_{\mathbf{jl}}\left(  \left\vert z^{\prime}\right\vert
\right)  f_{\mathbf{l}}^{+}\left(  z_{>}\right)  f_{\mathbf{l}}^{-}\left(
z_{<}\right)  ,\nonumber
\end{align}
Similarly, the form of thematrix elements for $\bar{D}_{cc}\left(
z,z^{\prime}\right)  $ and $\bar{D}_{oc}\left(  z,z^{\prime}\right)  $ are of
the same given by%
\begin{align}
\bar{D}_{\mathbf{jl}}\left(  z^{\prime},z\right)  =  &  \delta_{\mathbf{jl}%
}\delta_{z^{\prime}z}\label{Eq:Dccmatelem}\\
&  +e^{i\left(  \phi_{K}^{\mathbf{l}}-\phi_{K}^{\mathbf{j}}\right)  z^{\prime
}}U_{\mathbf{jl}}\left(  \left\vert z^{\prime}\right\vert \right)
\dfrac{\alpha_{\mathbf{l}}^{\left\vert z-z^{\prime}\right\vert +1}}{J_{K}%
^{\mu\nu}\left(  1-\alpha_{\mathbf{l}}^{2}\right)  },\nonumber
\end{align}
where $\alpha_{\mathbf{l}}$ is given by Eq. \ref{Eq:closedbandgf}.

\begin{widetext}

Inverting $\bar{D}$ directly gives%
\[
\bar{D}^{-1}=\left(
\begin{array}
[c]{cc}%
\left(  \bar{D}_{oo}-\bar{D}_{oc}\bar{D}_{cc}^{-1}\bar{D}_{co}\right)  ^{-1} &
-\bar{D}_{cc}^{-1}\bar{D}_{co}\left(  \bar{D}_{oo}-\bar{D}_{oc}\bar{D}%
_{cc}^{-1}\bar{D}_{co}\right)  ^{-1}\\
-\bar{D}_{oo}^{-1}\bar{D}_{oc}\left(  \bar{D}_{cc}-\bar{D}_{co}\bar{D}%
_{oo}^{-1}\bar{D}_{oc}\right)  ^{-1} & \left(  \bar{D}_{cc}-\bar{D}_{co}%
\bar{D}_{oo}^{-1}\bar{D}_{oc}\right)  ^{-1}%
\end{array}
\right)  .
\]
In Eq. \ref{Eq:AppKmat}, we are only concerned with the open-open segment.
Inserting $\bar{D}$ and carrying out the matrix multiplication gives%
\begin{align}
\mathbf{K}_{\mathbf{ij}}^{L}  &  =-2\pi\left\{  \sum_{z,z^{\prime}%
}f_{\mathbf{j}}^{+}\left(  z^{\prime}\right)  \left[  \left(  \bar{D}%
_{oo}-\bar{D}_{oc}\bar{D}_{cc}^{-1}\bar{D}_{co}\right)  ^{-1}\bar{U}%
_{oo}\right]  _{\mathbf{ij}}^{r}f_{\mathbf{i}}^{+}\left(  z\right)  \right.
\label{Eq:Kmatgenint}\\
&  -\left.  \sum_{z,z^{\prime}}f_{\mathbf{j}}^{+}\left(  z^{\prime}\right)
\left[  \bar{D}_{oo}^{-1}\bar{D}_{oc}\left(  \bar{D}_{cc}-\bar{D}_{co}\bar
{D}_{oo}^{-1}\bar{D}_{oc}\right)  ^{-1}\bar{U}_{co}\right]  _{\mathbf{ij}%
}f_{\mathbf{i}}^{+}\left(  z\right)  \right\}  .\nonumber
\end{align}
Where $\bar{U}_{\mathbf{ji}}=e^{i\left(  \phi_{K}^{\mathbf{i}}-\phi
_{K}^{\mathbf{j}}\right)  z}U_{\mathbf{ji}}\left(  \left\vert
z\right\vert \right)  $ are matrix elements of the interaction in the two-body
Wannier basis.

Equation \ref{Eq:Kmatgenint} is general for interactions of any range. Here,
we are concerned with on-site interactions where $\bar{U}_{\mathbf{ji}%
}=e^{i\left(  \phi_{K}^{\mathbf{i}}-\phi_{K}^{\mathbf{j}}\right)
z}U_{\mathbf{ji}}\delta_{z,0}$. Inserting this collapses the double
sum and we may evaluate this at $z=z^{\prime}=0$. We may also note that
$f_{\mathbf{l}}^{-}\left(  0\right)  =0$ while $f_{\mathbf{i}}^{+}\left(
0\right)  =\sqrt{\lambda/\pi v_{g}^{\mathbf{i}}}.$ This simplifies the
expression for the $D_{\mathbf{ji}}$ matrix elements considerably for open
channels leaving $\bar{D}_{oo}=\mathbb{1}$ and $D_{co}=\mathbb{0}$ Inserting this gives%

\begin{align}
\mathbf{K}_{\mathbf{ij}}^{L}=&-2\pi\left\{  \sum_{z,z^{\prime}%
}f_{\mathbf{j}}^{+}\left(  z^{\prime}\right)  \left[  \bar{U}_{oo}\right]
_{\mathbf{ij}}f_{\mathbf{i}}^{+}\left(  z\right)  
 - \sum_{z,z^{\prime}}f_{\mathbf{j}}^{+}\left(  z^{\prime}\right)
\left[  \bar{D}_{oc}\bar{D}_{cc}^{-1}\bar{U}_{co}\right]  _{\mathbf{ij}%
}f_{\mathbf{i}}^{+}\left(  z\right)  \right\}  ,\nonumber\\
=&-2\lambda\left(  v_{g}^{\mathbf{i}}%
v_{g}^{\mathbf{j}}\right)  ^{-1/2}\left[  \mathcal{U}_{oo}-\mathcal{U}%
_{oc}\bar{g}_{c}\left(  \mathbb{1}+\mathcal{U}_{cc}\bar{g}_{c}\right)
^{-1}\mathcal{U}_{co}\right]  _{\mathbf{ij}}\nonumber,
\end{align}
\end{widetext}
which is the expression that appears in Eq. \ref{Eq:Kmatfinal}.

\bibliography{AllRefs.bib}
\end{document}